\def\be{\begin{equation}}
\def\te{\end{equation}}
\def\bea{\begin{eqnarray}}
\def\nn{\nonumber}
\def\tea{\end{eqnarray}}

\def\ha{{1\over 2}}

\def\xib{\overline{\xi}}

\def\a{\alpha}
\def\b{\beta}
\def\d{\delta}
\def\e{\epsilon}

\def\k{\kappa}

\def\m{\mu}

\def\n{\nu}

\def\D{\Delta}

\def\L{\Lambda}
\def\O{\Omega}

\def\bb{\bibitem}
\def\mb{\overline{m}}

\documentstyle[aps]{revtex}
\input{epsf}

\begin{document}
\title{Vacuum Effects of Ultra-low Mass Particle Account for Recent
Acceleration of Universe}
\author{Leonard Parker\thanks{Electronic address:
leonard@uwm.edu} and Alpan Raval\thanks{Electronic address: raval@uwm.edu}\\
{\small Department of Physics, University of Wisconsin-Milwaukee, P.O. Box 
413,
 Milwaukee, WI 53201.}}
\maketitle
\begin{abstract}
In recent work, we showed that non-perturbative vacuum effects of a
very low mass particle could induce, at a redshift of order 1, a
transition from a matter-dominated to an accelerating universe. In
that work, we used the simplification of a sudden transition out of
the matter-dominated stage and were able to fit the Type Ia supernovae
(SNe-Ia) data points with a spatially-open universe. In the present
work, we find a more accurate, smooth {\it spatially-flat}  
analytic solution to the
quantum-corrected Einstein equations. This solution gives a good fit
to the SNe-Ia data with a particle mass parameter $m_h$ in the range
$6.40 \times 10^{-33}$ eV to $7.25 \times 10^{-33}$ eV. It follows that the
ratio of total matter density (including dark matter) to critical
density, $\O_0$, is in the range $0.58$ to $0.15$, and the age $t_0$ of
the universe is in the range $8.10\,h^{-1}$ Gyr to $12.2\,h^{-1}$ Gyr, where
$h$ is the present value of the Hubble constant, measured as a
fraction of the value $100$ km/(s Mpc). This spatially-flat model 
agrees with estimates of the
position of the first acoustic peak in the small angular scale
fluctuations of the cosmic background radiation, and with light-element
abundances of standard big-bang 
nucleosynthesis. 
Our model has only a single free
parameter, $m_h$, and does not require that we live at a special time
in the evolution of the universe.
WISC-MILW-99-TH-10
\end{abstract}
\newpage
\section{Introduction}

Many attempts \cite{sahni,bahcall,wang} have been made to account for 
the unexpected behavior of the recent expansion of the universe.
In a previous paper \cite{parrav2}, we showed that non-perturbative quantum
effects in the vacuum may account for the SNe-Ia data \cite{reiss,perl}
 suggesting
a recent acceleration of the expansion of the universe. We considered
a free quantized scalar field (with possible coupling to the 
scalar curvature, $R$). The propagator of the field included an infinite
sum of all terms having at least one factor of $R$
\cite{partom,jack}. 
The effective
action was given in \cite{parrav2}, which can be referred to by the reader for 
a more complete exposition of the theory. We found an approximate
solution to the effective Einstein gravitational field equations.
In that solution, there was a sudden transition (at redshift $z$ of
order $1$) from an earlier matter dominated stage of the expansion to a 
mildly inflating de Sitter expansion. This sudden transition model
required the universe to be spatially open in order to fit the
cosmological data.

Here we find an improved analytic solution to the effective Einstein 
equations and determine its consequences. Surprisingly, we find that
the analytic solution permits the SNe-Ia data to be fit with a 
spatially-flat model having a reasonable matter density and age. 
The spatial flatness
makes a significant difference \cite{kami} in attempting to fit the spectrum of
small angular scale fluctuations of the cosmic microwave background
radiation (CMBR). It appears that the existing data favors a
spatially-flat cosmological model. Spatial flatness is also favored in
models of the universe having an early inflationary stage.

In our present model, there is only one free parameter, namely, the 
ratio of the mass parameter $\mb$ to the present value
of the Hubble constant, $H_0$. It is convenient to express this
parameter in the form
\be
\label{mh}
m_h \equiv \mb/h,
\te
where 
\be
h \equiv H_0/(100\, {\rm km/(s\, Mpc)}),
\te
with $H_0$ expressed in km/(s Mpc).
We find that a good fit to the SNe-Ia data is obtained when $m_h$ is
in the range $6.40 \times 10^{-33}\,{\rm eV} < m_h < 7.25 \times
10^{-33}\,{\rm eV}$. For this parameter range, the ratio of the total
matter density (including dark matter) to critical density, $\O_0$, is
found to
be in the  range $0.58 > \O_0 > 0.15$, and the age $t_0$ of the
universe is found to be in the range $8.10h^{-1}\,{\rm Gyr} < t_0 <
12.2h^{-1}\,{\rm Gyr}$. Also, we find that our model gives
reasonable abundances for light elements formed during big-bang
nucleosynthesis. 
In our model, no special coincidence at the present time is necessary to
explain why the energy density of matter is of the same order
as the vacuum energy density.

The organization of this paper is as follows. In Section II, we
give the effective Einstein equation of the model, and summarize the
solution of Ref. \cite{parrav2}. In Section III, we derive an analytic
solution for the scale factor of the model. In Section IV, we compare
the model to SNe-Ia data and obtain ranges for $m_h$, $\O_0$ and
$t_0$. In Section V, we discuss the implications of the model for
big-bang nucleosynthesis and for the spectrum of CMBR fluctuations. In
Section VI, we show it is probable (independent of fitting the observations)
that, at the present time, the vacuum
energy density in our model is comparable to the matter density.
Finally, our conclusions are given in Section VII.    

\section{Our Model}

In Ref. \cite{parrav2}, we consider a free, massive quantized scalar field 
of inverse Compton wavelength (or mass) $m$, 
and curvature
coupling $\xi$. 
The effective action for gravity coupled to such a field is obtained by
integrating out the vacuum fluctuations of 
the field \cite{parrav2,partom}.
This effective action is 
the simplest one that gives
the standard trace anomaly in the massless-conformally-coupled limit, and
contains the explicit sum (in arbitrary dimensions) of all terms in the 
propagator having at
least one factor of the  scalar curvature, $R$. 

The trace of the Einstein equations obtained by variation of
this effective action with respect to the metric tensor take the
following form in a
Friedmann-Robertson-Walker (FRW) spacetime (in units such that $c=1$):
\bea
\label{one}
R + \frac{T_{cl}}{2\k_o} -4 \L _o &=& \frac{\hbar m^2}{32 \pi^2
\k_o}\left\{\vphantom{\frac{m^2}{M^2}} (m^2 +\xib R)\ln 
\mid 1+\xib R m^{-2}\mid \right.\nn \\
& &-\left. \frac{m^2\xib
R}{m^2 +\xib R}\left(1+{3\over 2}\xib \frac{R}{m^{2}} + \ha \xib^2 
\frac{R^2}{m^{4}} (\xib^2 - (1080)^{-1}) +v \right)\right\}, 
\tea
where $T_{cl}$ is the trace of the stress tensor of a classical, 
perfect fluid component containing mixed matter and radiation, 
$\L _{o}$ is the cosmological constant, 
$\k_o = (16\pi G)^{-1}$
($G$ is Newton's constant), $\xib = \xi -1/6$, and $v$ is a quantity
that vanishes in de Sitter space:
\be
v = \frac{1}{180m^4}\left({1\over 4}R^2 - R_{\m \n}R^{\m \n}\right).
\te
Here, the  metric is
\be
ds^2 = -dt^2 + a(t)^2\left(\frac{dr^2}{1-kr^2} + r^2 d\O^2\right), 
\te
where $a(t)$ is the scale factor and $k =0, 1, -1$ give spatially
flat, closed and open universes respectively.  

The right-hand-side (RHS) of Eq. (\ref{one}) is proportional to the trace of
the quantum contribution to the stress tensor. In 
Eq. (\ref{one}), we have assumed that terms
in the quantum stress tensor that depend on derivatives of the
curvature are negligible. 
This assumption is consistent with the
solution that we obtain. 

In \cite{parrav2}, we found that Eq. (\ref{one}) admits a solution in
which a matter-dominated FRW universe transits to a stage of the
expansion in which the scalar curvature is nearly constant. 
This led us to construct an approximate cosmological
model which makes a {\it sudden} transition to a de Sitter universe out of a
matter-dominated one. In that approximation, we needed a spatially-open 
cosmology in order to fit high-redshift SNe-Ia data, 
and to get a reasonable value of the present
matter density. 

In this paper, we eliminate the assumption of a sudden transition to a
de Sitter expansion, by finding an analytic solution to the effective
Einstein equations of our model. Consistency with cosmological
observations requires that this improved analytic solution has zero
(or nearly zero) spatial curvature. This solution consists of a
matter-dominated universe smoothly, with continous first and second
derivatives of the scale factor (i.e., $C^2$), 
joined to a constant scalar curvature
expansion that asymptotes to a de Sitter universe at late times.
The  spatial flatness is consistent with 
estimates of the first acoustic peak of small angular scale
fluctuations in the CMBR,
which suggest that the universe is spatially flat (although this issue
awaits an ultimate decision at the time of writing, see
Ref. \cite{turner} for details). 

We now turn to the derivation of a smooth
solution for the cosmological scale factor, $a(t)$.  

\section{Analytic Solution}

In the following analysis, as in \cite{parrav2}, we take $\L_o = 0$. 
For $\xib<0$, and with a low 
value of  
$\overline{m}$ ($\simeq 10^{-33}$ eV), 
we find in \cite{parrav2} that the evolution of the
FRW universe is essentially unaffected by the quantum contributions at
early times. A
perturbative analysis in $\hbar$ shows that quantum
contributions to the stress tensor begin to have a significant effect 
at a time $t_j$, 
when the classical
scalar curvature has decreased to a value given roughly by 
\be
\label{two}
R_{cl}(t_j) \equiv \frac{\rho_m(t_j)}{2\k_o} =
\overline{m}^2,
\te
where
\be
\mb^2 \equiv m^2/(-\xib).
\te
This effect is shown to occur well into the matter-dominated stage
of the evolution (therefore, for $t>t_j$, $T_{cl} \simeq -\rho_m$). 
We further argue that, after the scalar curvature
reaches the value $R_{cl}$ (i.e. for $t\geq t_j$), 
it stays essentially constant during the
later evolution of the universe. 
More
precisely, for all times $t>t_j$, we find that the scalar curvature 
has the form
\be
\label{r}
R(t) = \mb^2(1-\e(t)),
\te 
where
\be
\label{ep}
\e(t) = \ha \xib \left(-\d + (\d^2 + 4\beta \xib^{-1})^{1/2}\right),
\te
and
\be
\label{del}
\d = -\frac{\rho_m(t)}{2m^2\k_o} - \xib^{-1},~~~~\b =
 \frac{r}{2\pi}\left(v(t)- \frac{1}{2160\xib^2}\right).
\te
Here,
\be
r = \frac{m^2}{m_{Pl}^2},
\te
$m_{Pl}$ being the Planck mass.
As shown in \cite{parrav2}, it is straightforward to verify 
that $\e (t)$ is of order $r$ for $t>t_j$
and of order $\sqrt{r}$ at $t=t_j$. For the low value of $r$ under
consideration,  it follows that  
$\e (t) \ll 1$ for all $t \geq t_j$. Thus the scalar curvature stays
nearly constant at the value $\mb^2$ for $t \geq t_j$. 

Despite the constancy of the scalar curvature,
the quantum contribution to the stress tensor increases
dramatically from the time $t_j$ to the present time $t_0$. To see
this, consider the trace of the quantum stress-tensor, $T_q$,
defined by $T_q/(2\k_o) = -($RHS of
(\ref{one})$)$. It follows from (\ref{one}) that (with $\L_o =0$),
\bea
\label{quantt}
T_q &=& -2\k_o R - T_{cl} \nn \\
&=& -2\k_o \mb^2 + \rho_m + {\cal O}(\e ),
\tea
where the second approximate equality holds for $t \geq t_j$. Assuming
(as we find)
that the transition time $t_j$ occurs at a redshift $z_j$ of order $1$, and
given that $\rho_m (t_j) \simeq 2\k_o \mb^2$ (from Eq. (\ref{two})),
we obtain the matter density at the present time, $\rho_m (t_0)$, as
\bea
\rho_m (t_0) = -T_{cl}(t_0)&=& \left(a(t_j)/a(t_0)\right)^3
\rho_m (t_j)  
\simeq (1+z_j)^{-3}2\k_o \mb^2 \nn \\
&\simeq& \k_o \mb^2/4 + {\cal O}(\e ).
\tea
Thus, according to
Eq. (\ref{quantt}), $T_q$ grows from $T_q(t_j) \simeq
-2\k_o \mb^2 + 2\k_o \mb^2 \simeq 0$ at time $t_j$ to $T_q(t_0) \simeq
-2\k_o \mb^2 + (1/4)\k_o \mb^2 \simeq -(7/4)\k_o \mb^2 \simeq 7T_{cl}(t_0)$ at
the present time $t_0$. Thus, we find that the quantum contribution
to the stress tensor grows from a negligible value at a redshift of
order $1$ to a value exceeding the classical contribution at the
present time. 

To obtain the scale factor $a(t)$, we consider a spatially flat
($k=0$) cosmology, and find the constant scalar
curvature solution for $t \geq t_j$ that joins to the usual 
matter-dominated solution
at time $t_j$, with continous first and second derivatives. The time
$t_j$  obtained from Eq. (\ref{two}), using $R_{cl}(t) = (4/3)t^{-2}$
for
$t \leq t_j$ in the
spatially flat matter-dominated universe, is
\be
t_j = (2/\sqrt{3})\,\mb^{-1}. 
\te
For such a universe the Hubble constant $H(t) = (2/3)t^{-1}$, 
has the value at $t_j$ given by
\be
\label{htj}
H(t_j) = \mb/\sqrt{3}.
\te
For $t>t_j$, we now obtain the unique solution to the constant scalar 
curvature equation
\be
\label{coneq}
R = 6\left(\frac{\ddot{a}}{a} + \frac{\dot{a}^2}{a^2}\right) = \mb^2,
\te
that satisfies
\be
H(t_j) = \dot{a}(t_j)/a(t_j) = \mb/\sqrt{3}.
\te
Continuity of the scalar
curvature at $t_j$ ensures that the second derivative of the scale factor is
also continuous. This solution is
\be
\label{met1}
a(t) = a(t_j)
\sqrt{\sinh\left(\frac{t\mb}{\sqrt{3}}-\a\right)\left/\sinh\left(
{2\over 3} - \a\right)\right.},~~~t>t_j,
\te
where
\be
\label{c}
\a = \frac{2}{3}- \tanh^{-1}\left({1\over 2}\right) \simeq 0.117.
\te
We have verified that this solution also satisfies the remaining
Einstein equations up to terms of order $\e $.
According to Eq. (\ref{met1}), the expansion approaches a 
de Sitter expansion at late times
(i.e., $t\mb \gg 1$). Furthermore, it has the property that the
deceleration parameter $q_0 \equiv -\ddot{a}a/\dot{a}^2$ is negative (i.e., the
universe is accelerating) for all
$t>t_a$, where
\be
t_a = \sqrt{3}\,\mb^{-1}\left(\a + \tanh^{-1}(2^{-1/2})\right) \simeq
1.73\, \mb^{-1} \simeq 1.50\, t_j.
\te
Also, the solution joins in a smooth ($C^2$) manner to the usual
spatially flat matter-dominated solution for $t<t_j$, i.e.
\be
\label{met2}
a(t) = a(t_j)\left(t/t_j\right)^{2/3} =
\left(\sqrt{3}\mb t/2\right)^{2/3},~~~t<t_j.
\te

This solution, given by Eqs. (\ref{met1}), (\ref{c}) and (\ref{met2}),
depends on only one parameter, $\mb$. It is more accurate than the
approximation of a sudden transition to a de Sitter expansion, that we
used in \cite{parrav2}. Our analytic solution takes a long time (of
order $\mb^{-1}$) to approach a de Sitter expansion. This is the
reason that the spatially flat model gives agreement with observations
(as we show below). In the approximation of a sudden transition to de
Sitter, consistency with observations required a negative spatial
curvature. We further note that our analytic solution is not the same as that
of a mixed matter-cosmological constant model, for which $a(t)
\propto (\sinh(\sqrt{3\L}\,t/2))^{2/3}$.

Further  improvements in the accuracy of our solution can be made by
an iterative procedure in which the present solution is used to
evaluate $\e (t)$ in Eqs. (\ref{ep}) and (\ref{del}), and then
Eq. (\ref{coneq}) is solved with $\mb^2$ replaced by the RHS of Eq.
(\ref{r}).

\section{Comparison with Observations}

For comparison with observations of high-redshift Type 1a supernovae
(SNe-Ia), we calculate the luminosity distance-redshift relation for
the model defined by Eqs. (\ref{met1}) and (\ref{met2}), 
and, from it, the difference in
apparent and absolute magnitudes as a function of the redshift $z$ of
a source. This difference is given by 
\be
\label{diffmag}
m-M = 5 \log_{10}d_L  + 25,
\te
where $d_L$ is the luminosity distance to the source in Mpc, 
defined as \cite{wein}
\be
\label{lum1}
d_L = (1+z)a_0 r_1,
\te
where $a_0$ is the present scale factor, and $r_1$ is the comoving
coordinate distance from a source at redshift $z$ to a detector at
redshift $0$. 
For a spatially flat FRW universe, $r_1$ is given by 
\be
\label{lum3}
r_1 = ca_0^{-1}\int_0^z \frac{dz'}{H(z')},
\te
where $H(z)$ is the Hubble constant as a function of $z$, and the
speed of light $c$ is now shown explicitly. It is
convenient to introduce the parameter $h \equiv H_0/(100$ km/(s Mpc)), 
where $H_0$
is the present value of the Hubble constant in km/(s Mpc). 
It is also useful to work with the rescaled mass parameter $m_h$ of
Eq. (\ref{mh}).
We note that $m_h$, being an inverse length, can be measured in units
of Mpc$^{-1}$.

For our model, we define $z_j$ as the
redshift at time $t_j$. Then we obtain, using Eqs. (\ref{met1}) and
(\ref{lum3}), 
\be
\label{rol}
hr_{1<}(z) = a_0^{-1}\sqrt{12}m_h^{-1}\chi
\int_{1}^{1+z}\frac{dx}{\sqrt{x^4 + \chi^2}},
\te
where $r_{1<}(z)$ denotes $r_1(z)$ for $z <z_j$, and
\be
\chi = \sinh \left(\frac{cht_0\, m_h}{\sqrt{3}} - \a\right).
\te
For $z >z_j$, we use Eqs. (\ref{met2}) and (\ref{lum3}) to obtain
\be
\label{rog}
hr_{1>}(z) = hr_{1<}(z_j) + 2\sqrt{3}
a_0^{-1}m_h^{-1}(1+z_j)^{3/2}\left( (1+z_j)^{-1/2} - (1+z)^{-1/2}\right).
\te
Eq. (\ref{lum1}) gives the luminosity distance for this model, 
$d_{L1}(z)$, defined by
\bea
\label{lumd1}
d_{L1}(z) &=& (1+z)a_0 r_{1<}(z), ~~~z<z_j \nn \\
&=& (1+z)a_0 r_{1>}(z), ~~~z>z_j. 
\tea
After substituting for $r_{1<}(z)$ and $r_{1>}(z)$ in the above
equations, $hd_{L1}(z)$ does not depend on $a_0$ (for a spatially flat
universe).  The three parameters that
occur in $hd_{L1}$ are $ht_0$, $z_j$ and $m_h$. However, we may
differentiate 
Eq. (\ref{met1}) to obtain
\be
\label{phubble}
H(t) = \frac{c\mb}{\sqrt{12}}\coth\left(\frac{ct \mb}{\sqrt{3}} -
\a\right),
\te
The above equation, evaluated at the present cosmic time $t_0$, can be
rewritten in the form
\be
\label{ageexp}
ht_0 = (3.26 \times 10^6 \, {\rm yr/Mpc})
m_h^{-1}\left(\tanh^{-1}\left((865.4\, {\rm Mpc})m_h\right) +\a\right),
\te
which gives $ht_0$ in years.
Also, Eq. (\ref{met1}) yields
\be
\label{zj}
1+z_j \equiv a(t_0)/a(t_j) = \sqrt{\sinh\left( 
\frac{cht_0\, m_h}{\sqrt{3}}-\a\right)\left/\sinh\left(
{2\over 3} - \a\right)\right.}.
\te
Eq. (\ref{ageexp}) shows that $ht_0$ is a function of
$m_h$ alone. Therefore, the redshift $z_j$ as given by Eq. (\ref{zj}) is
also a function of $m_h$ alone. This implies that the function
$hd_{L1}(z)$ depends on a single parameter, $m_h$.
The present ratio of the matter density
to critical density, $\O_0$, is also a function of the same parameter. 
To see this, we use $\rho_m \propto a^{-3}$. Then one has 
\be
\O_0 \equiv \frac{8\pi G}{3H_0^2} \rho_{m}(t_0) = 
\frac{8\pi G}{3H(t_j)^2} 
\rho_{m}(t_j) 
\left(\frac{H(t_j)}{H_0}\right)^2
\left(\frac{a(t_j)}{a_0}\right)^3.
\te
Continuity with the spatially flat matter-dominated universe at
$t=t_j$ requires that \\$(8\pi G \rho_m(t_j))/(3H(t_j)^2) =
1$. Therefore
\bea 
\label{omo}
\O_0 &=& \left(H(t_j)/H_0\right)^2
\left(a(t_j)/a_0\right)^3 \nn \\
&=& (2.996 \times 10^6\,{\rm Mpc}^2) m_h^2 \left(\sinh\left(
\frac{cht_0\,m_h}{\sqrt{3}}-\a\right)\left/\sinh\left(
{2\over 3} - \a\right)\right.\right)^{-3/2},
\tea
where we have used Eqs. (\ref{htj}) and (\ref{met1}) in arriving at
the last equality. Note that the quantity $ht_0$ appearing in the above
equation is itself a function of $m_h$ (see Eq. (\ref{ageexp})). 
Therefore $\O_0$ is also a function of the parameter $m_h$,
which is
the only adjustable parameter in our model.

If the time $t_j$ is less than the present age $t_0$,
the monotonic behavior of Eq. (\ref{phubble}) 
implies that $H(\infty)< H_0 < H(t_j)$. With $H(\infty)$ and $H(t_j)$
obtained from
Eq.(\ref{phubble}), the previous inequality gives
\bea
\label{ineq1}
\mb &>& \sqrt{3}H_0/c = 5.78 \times 10^{-4} h\,\, {\rm Mpc}^{-1},\\
\mb &<& \sqrt{12}H_0/c = 1.16 \times 10^{-3} h\,\, {\rm Mpc}^{-1}.
\tea
We therefore find that the rescaled mass parameter $m_h$ is constrained 
by the model 
to lie in the range
\be
\label{mrange1}
5.78 \times 10^{-4}\, {\rm Mpc}^{-1} < m_h < 1.16 \times 10^{-3}\, {\rm Mpc}^{-1}.
\te
The above range on $m_h$ can be also expressed in electron-volts
(eV), as
\be
\label{mrange}
3.69 \times 10^{-33}\, {\rm eV} < m_h < 7.39 \times 10^{-33}\,{\rm
eV}.
\te
We emphasize that the above range of values of $m_h$ is a
{\it consequence} of our model, if $t_0 >t_j$, 
independent of any fit to cosmological
observations.

\begin{figure}[hbt]
\centering
\leavevmode
\epsfysize=6.in\epsffile{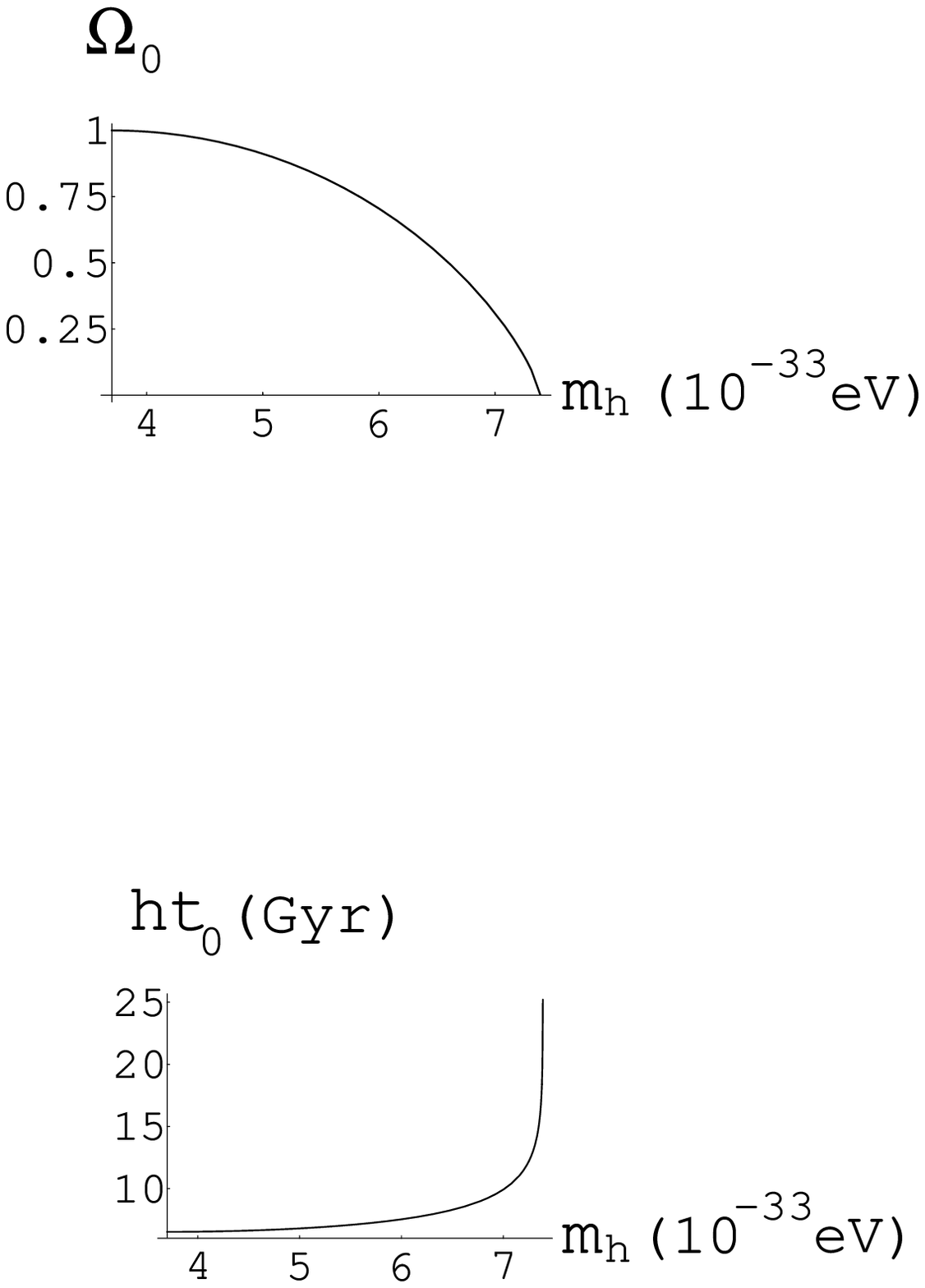}
\caption{Plots of $\O_0$ and $ht_0$ versus $m_h$, 
with the range of $m_h$ values
$3.69 \times 10^{-33}$ eV $<m_h<$ $7.40 \times 10^{-33}$ eV.}   
\label{figu3}
\end{figure}

\begin{figure}[hbt]
\centering
\leavevmode
\epsfysize=6.in\epsffile{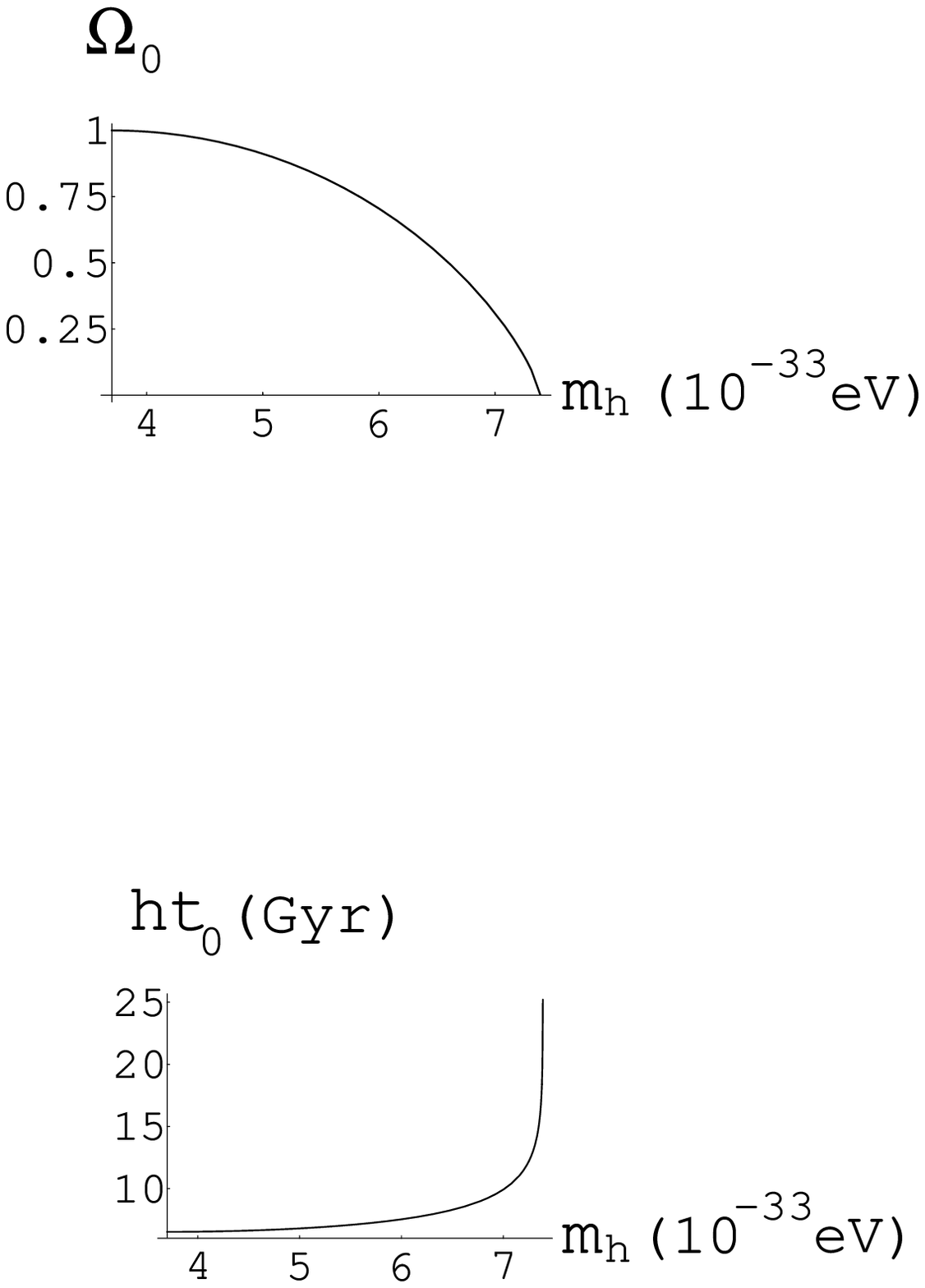}
\caption{A plot of $\O_0$ versus $ht_0$, with the range of $m_h$ values
$3.69 \times 10^{-33}$ eV $<m_h<$ $7.40 \times 10^{-33}$ eV.}   
\label{figu4}
\end{figure}

Plots of $\O_0$ and $ht_0$ vs. $m_h$, using Eqs. (\ref{omo}) and
(\ref{ageexp}) respectively, are given in Fig. \ref{figu3}, for the
range of $m_h$ values of Eq. (\ref{mrange}) above. For the same range
of $m_h$ values, Fig. \ref{figu4} is a plot of $\O_0$ vs. $ht_0$.

We now use the luminosity distance of Eq. (\ref{lumd1}) to fit the
SNe-Ia data. To do so, it is convenient to normalize the difference
$m-M$ of Eq. (\ref{diffmag}) to its value in an open universe with
$\O_0 =0.2$. We define
\be
\label{diffmnorm}
\D (m-M)(z) \equiv 5 \log_{10}\left(\frac{d_{L1}(z)}{d_{L2}(0.2,
z)}\right) = 5 \log_{10}\left(\frac{hd_{L1}(z)}{hd_{L2}(0.2, z)}\right),
\te 
where
\bea
d_{L2}(\O_0, z) &=& 2H_0^{-1}c\,\O_0^{-2}\left(\O_0 z +(\O_0-2)
\left(\sqrt{1+\O_0 z} -1\right)\right)\nn \\
&=& h^{-1}(5995.8 \, {\rm Mpc})\,\O_0^{-2}\left(\O_0 z +(\O_0-2)
\left(\sqrt{1+\O_0 z} -1\right)\right).
\tea
Since $hd_{L1}(z)$ is a function of $m_h$ and $z$, and $hd_{L2}(0.2, z)$
is a function of $z$ alone, it follows that $\D (m-M)$ is a function
of $m_h$ and $z$. 

\begin{figure}[hbt]
\centering
\leavevmode
\epsfysize=5.in\epsffile{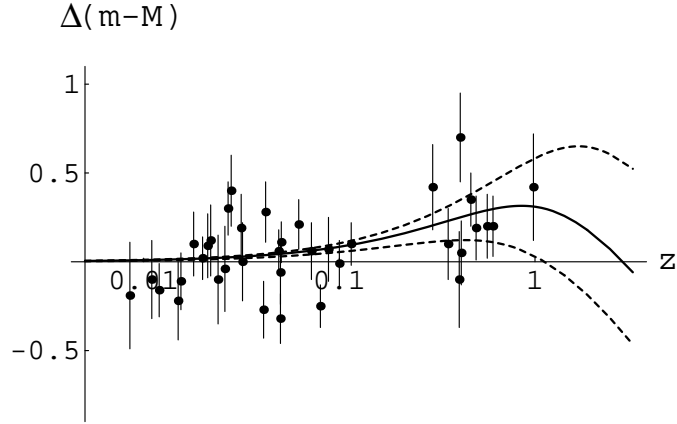}
\caption{A plot of the difference between apparent and absolute
magnitudes, 
as functions of redshift $z$, normalized to an open universe with
$\O_0 =0.2$ and zero cosmological constant. The points with vertical
error bars represent SNe-Ia data obtained from
Ref.[5]. 
The two
dashed curves represent the values (a) $m_h = 6.40 \times
10^{-33}$ eV (lower dashed curve), and (b) $m_h
= 7.25 \times 10^{-33}$ eV (upper dashed curve). 
The solid curve 
represents the intermediate value (c) $m_h = 6.93 \times 10^{-33}$ eV.}
\label{figu1}
\end{figure}

Fig. \ref{figu1} is a plot of $\D(m-M)$ vs. $z$, along with a plot of
SNe-Ia data acquired from Ref. \cite{reiss}. A good fit is obtained by
any curve that lies in between the two  dashed curves shown, where curve (a)
has 
$m_h = 6.40 \times 10^{-33}$ eV,
and curve (b) has $ m_h = 7.25 \times 10^{-33}$ eV. This gives a value
of $\O_0$ in the range 
\be
0.15 < \O_0 < 0.58,
\te
and a value of $ht_0$ in the range
\be
\label{htz2}
8.10\, {\rm Gyr} < ht_0  < 12.2\, {\rm Gyr}.
\te
Estimates of $h$ give $0.55 < h < 0.75$ \cite{bahcall}. For $h=0.65$, 
Eq. (\ref{htz2}) gives a
range for the age of the universe, namely
\be
12.5\, {\rm Gyr} < t_0 < 18.8\, {\rm Gyr}.
\te

\begin{figure}[hbt]
\centering
\leavevmode
\epsfysize=6.in\epsffile{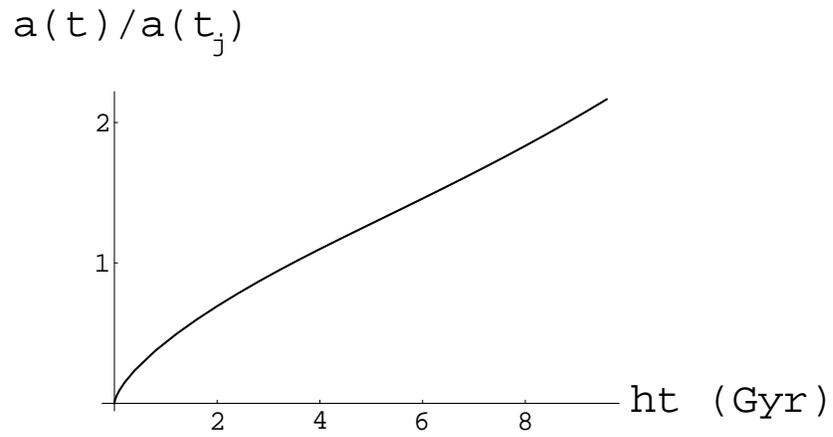}
\caption{A plot of $a(t)/a(t_j)$ versus $ht$ in our  
model universe, for the value $m_h = 6.93 \times 10^{-33}$ eV. 
The graph terminates at the $x$-coordinate 
value $ht_0$, $t_0$ being the present age of the universe.}
\label{figu2}
\end{figure}

A representative solid curve (c) has the value $m_h = 6.93 \times
10^{-33}$ eV, which gives $\O_0 = 0.346$ and $t_0 = 14.8$ Gyr (with $h=0.65$).
A plot of the scale factor $a(t)$ for this curve is shown in
Fig. \ref{figu2}.

\section{Recombination, Nucleosynthesis and CMBR fluctuations}

In this section, we first calculate the time of recombination in our model,
compare it to the recombination time in a standard open universe, and
discuss the implications for big-bang nucleosynthesis. We then show that
the apparent angular size of CMBR fluctuations in our model is somewhat 
smaller than in a spatially flat $\O_0 =1$ model, although consistent
with available data.

In our
model, we find that recombination occurs in the matter-dominated
era. The redshift at recombination is given by the following expression,
discussed in Ref. \cite{hu},
\be
\label{hurecomb}
z_r = 1048 \left(1+0.0124 (\O_b h^2)^{-0.738}\right)\left(1+g_1(\O_0
h^2)^{g_2} \right),
\te
where
\bea
g_1 &=& 0.0783 (\O _b h^2)^{-0.238} \left(1+39.5 (\O _b
h^2)^{0.763}\right)^{-1}, \\
g_2 &=& 0.560 \left(1+21.1 (\O _b h^2)^{1.81}\right)^{-1},
\tea
and $\O _b$ is the
ratio of  baryon density and critical density at the present time.  
For the ranges $0.0025 < \O _b h^2 < 0.25$, $0.15<\O_0<0.60$ and 
$0.55<h<0.75$ \cite{bahcall}, we find,
\be
\label{zr}
1055.73 < z_r < 1301.80.  
\te
The redshift at matter-radiation equality, $z_{eq}$, is given by
\be
1+z_{eq} = \O_0 h^2/(\O_r h^2),
\te
where $\O_r$ is the ratio of radiation energy density to critical density 
at the present time. With a CMBR temperature of $2.726$ K at the present time
\cite{mather}, we obtain $\O_r h^2 = 4.16 \times 10^{-5}$. Using the same
range of values as above for $\O_0$ and $h$, we find
\be
1090.14 < z_{eq} < 8115.90.
\te
Therefore, in our model, recombination occurs in the matter-dominated era.

We now invert Eqs. (\ref{met1})  and (\ref{met2}) to  obtain a
 range for the time of recombination, $t_r$, corresponding to the
range of $z_r$
in Eq. (\ref{zr}). A lower bound on $t_r$
is obtained with the higher value of $z_r$ and the lowest value of
$m_h$ consistent with a fit to the SNe-Ia data, i.e., 
$m_h = 6.40 \times 10^{-33}$ eV,
and an upper bound is obtained with the lower value of $z_r$ and the
highest value of $m_h$ consistent with the same data, i.e., $m_h =
7.25 \times 10^{-33}$ eV. 
We find
\be
\label{recor}
1.82 \times 10^5 {\rm years} < ht_r < 4.97 \times 10^5 {\rm years}.
\te
We compare the above range with the corresponding range of $h t_r$ in
a standard,
spatially open matter-dominated model, with the same ranges of $\O_0$,
$h$ and $\O_b h^2$ (and therefore the same range of $z_r$). The scale
factor for an open matter-dominated
model \cite{wein} leads to the parametrized equations
\bea
1+z_r &=& \frac{2(1-\O_0)}{\O_0}\left(\cosh \psi_r -1\right)^{-1},\nn \\
ht_r &=& (4.89 \times 10^9\,{\rm yr})\frac{\O_0}{(1-\O_0)^{3/2}}
\left(\sinh \psi_r -\psi_r\right),  
\tea
$\psi_r$ being a dimensionless parameter. The above equations give the
following 
range for $ht_r$ in an open matter-dominated universe,
\be
\label{recos}
1.79 \times 10^5 {\rm years} < ht_r < 4.89 \times 10^5 {\rm years}.
\te

Comparison of Eqs. (\ref{recor}) and (\ref{recos}) reveals that the
range of recombination times for the two models differ by only about
$1.7$ percent. Furthermore,
the scale factors prior to recombination are almost identical in both models
(the effect of spatial curvature in the open model is negligible for
times prior to the recombination time). Therefore, it is expected that
nuclear reaction calculations in big-bang nucleosynthesis would give
nearly identical results for the abundances of light elements in both
models. Since these calculations are known to give results in
agreement with observation for the standard open model,
nucleosynthesis calculations in our model would also give results that
agree with observations, for the same values of $\O_0$, $h$ and $\O_bh^2$.

We now turn to the calculation of the apparent angular size of a
fluctuation of a fixed proper size $D$ at the surface of last
scattering. This apparent angular size in our model 
will be compared to the apparent
angular size of the same fluctuation in a spatially flat FRW model
having $\O_0 = 1$, since the latter model is known to yield a CMBR fluctuation
spectrum consistent with data.

For simplicity, we  take the representative values $m_h = 6.93
\times 10^{-33}$ eV, $h = 0.65$, and
$\O_bh^2 =0.025$ in the discussion that follows. The
chosen value of $m_h$ corresponds to the intermediate curve (c) of 
Fig. \ref{figu1},
and gives $\O_0 = 0.346$. Eq. (\ref{hurecomb}) then gives the redshift
at last scattering, as $z_r = 1089$. 

The apparent angular size $\theta$ of a fluctuation of fixed proper
size $D$ at
last scattering is given by
\be
\label{the}
\theta = D/(a(t_r) r_1) = D (1+z_r)^2/d_{L1}(z_r),
\te
where $d_{L1}(z_r)$ is found from Eq. (\ref{lumd1}), which also gives
\be
hd_{L1}(1089) = 1.017 \times 10^7\, {\rm Mpc}.
\te
Therefore, with $h=0.65$,
\be
\label{theo}
\theta = (7.594 \times 10^{-2}\, {\rm Mpc}^{-1})D
\te
in our model.

In a spatially flat $\O_0 =1$ model, taking $h=0.65$ and $\O_bh^2 =
0.025$, Eq. (\ref{hurecomb}) gives $z_r = 1105$. 

The luminosity
distance $d_L$ for a spatially flat $\O_0 =1$ cosmology is given by
\be
hd_L(z) = (5995.8\,{\rm Mpc})\,(1+z)\left(1-(1+z)^{-1/2}\right),
\te
which gives
\be
hd_L(1105) = 6.433 \times 10^6 \, {\rm Mpc}.
\te   
Therefore, with $h=0.65$, and using Eq. (\ref{the}), we obtain
\be 
\label{thef}
\theta = (1.236 \times 10^{-1}\, {\rm Mpc}^{-1})D.
\te

Comparison of Eqs. (\ref{theo}) and (\ref{thef}) reveals that the
apparent angular size of a fluctuation of a given proper size at last
scattering is about $1.63$ times less in our model than in a spatially
flat $\O_0 =1$ model. If the first acoustic peak in the small angular scale
CMBR spectrum arises from a fluctuation of fixed proper size in all
models, then the above result implies that, in our model, this peak 
would be shifted to a
higher mode number relative to that in a $\O_0 =1$ spatially flat
model. In the latter model, the first peak is known to occur at mode
number $l \simeq 200$, and therefore it would occur at $l \simeq 326$
in our model. These two possibilities are both consistent with the
existing data on small angular scale CMBR fluctuations (see, for
example, \cite{bahcall} and references therein).         

\section{The Question of Fine-Tuning}

Recent observations of Type-Ia supernovae evidently indicate that the universe
has been in an accelerating phase from a redshift of order $1$ up to
the present time, which implies that the contribution of
matter to the total energy density is of the same order
of magnitude as the contribution of vacuum energy density at the
present time (in spatially flat models, this means that
$\O_0/(1-\O_0)$ is of order $1$). 
Why  should this be the case? In mixed matter and cosmological constant
models, this question requires  an explanation of why the
cosmological constant must be fine-tuned to a precise,
non-zero value. As pointed out in \cite{parrav2}, our model is
relatively insensitive to the value of the cosmological constant term.
However, in our model, it would appear
that, within its allowed range of Eq. (\ref{mrange}), the  parameter 
$m_h$ must be finely tuned to give values for
$z_j$ and
$\O_0/(1-\O_0)$ that are within an order of magnitude of $1$. 
It should be noted that
the allowed range of values of $m_h$ given by Eq. (\ref{mrange}) does
not by itself constrain $z_j$ and $\O_0$. The lowest allowed value of
$m_h$ gives $z_j =0$ and $\O_0 = 1$, and the highest allowed value
gives $z_j = \infty$ and $\O_0 = 0$.

Here, we argue that, in our model, values within an oder of magnitude 
of $1$ 
for $z_j$ and $\O_0/(1-\O_0)$ are more likely than other values. 
The argument that follows rests on two
assumptions: (i) $t_0 > t_j$, and (ii) all values of $m_h$ within the
allowed range given by Eq. (\ref{mrange}) have equal {\it a priori}
probability. Assumption (i) implies the range of $m_h$ values of
Eq. (\ref{mrange}), and assumption (ii) is reasonable in lieu of a detailed
fundamental theory that predicts the value of $m_h$.       

Given these two assumptions, one may compute the probability
distributions for $z_j$ and $\O_0$ in a straightforward manner, since
both quantities are functions of $m_h$ alone. By assumption (ii), the
probability distribution function for $m_h$, $P(m_h)$, has the form
\bea
P(m_h) &=& P_0,~~~~3.69 \times 10^{-33} {\rm eV} < m_h < 7.39 \times
10^{-33}  {\rm eV}, \nn \\
&=& 0,~~~~{\rm otherwise}.
\tea 
Normalization of $P(m_h)$ yields $P_0 = 2.70 \times 10^{32}/$
eV.
We then obtain the probability distribution functions $P(z_j)$ and
$P(\O_0/(1-\O_0))$ for $z_j$ and $\O_0/(1-\O_0)$ respectively, as
\bea
P(z_j) &=& P_0/ \mid\frac{dz_j}{dm_h}\mid,\\
P(\O_0/(1-\O_0)) &=& P_0/ \mid\frac{d(\O_0/(1-\O_0))}{dm_h}\mid,
\tea
where $z_j(m_h)$ and $\O_0 (m_h)$ are given by Eqs. (\ref{zj}) and
(\ref{omo}) respectively. We now compute the probability that $z_j$
and $\O_0/(1-\O_0)$ lie between $0.1$ and $10$ (i.e., they are 
within an order of magnitude of $1$). We find
\bea
P[0.1<z_j<10] &=& \int_{0.1}^{10}dz_j\, P(z_j) = 0.851, \nn \\
P[0.1<\O_0/(1-\O_0)<10] &=&
\int_{0.1}^{10}d\left(\frac{\O_0}{1-\O_0}\right)\, 
P\left(\frac{\O_0}{1-\O_0}\right) = 0.619. 
\tea

It is therefore more likely for $z_j$ and $\O_0/(1-\O_0)$ to be within an
order of magnitude of $1$ rather than outside that range, 
assuming that all allowed values of $m_h$
have equal {\it a priori} probability.

\section{Conclusions}

We have shown that vacuum effects of a free scalar field of very low
mass can account for the observed acceleration (i.e., the SNe-Ia data), 
while at the same time predicting reasonable values for the age 
of the universe and the total matter density. Evidently, 
our model also predicts reasonable light element abundances, 
and as a consequence of spatial flatness is in agreement with 
current data on small angular scale CMBR fluctuations. 

Better SNe-Ia data would be able to distinguish between our model and
mixed matter-cosmological constant models, as well as quintessence
models \cite{wang}. 
The curves of $\D (m-M)$ vs $z$ are different in our model
from those of the other models. Future observations of small angular size 
CMBR fluctuations may also distinguish between these models.

We emphasize that our model is based on a free renormalizable quantum 
field and does not require that we live at a very special time in 
the evolution of the universe. We also note that a 
graviton field of very low mass may give rise to vacuum
effects similar to those of the scalar field we considered here.
In contrast, a scalar particle of very high mass with similar
characteristics to the present one, may contribute to a stage
of early inflation, with  reheating and exit from inflation
caused by particle production from the vacuum and possibly from
other inflationary exit mechanisms.\\

\bigskip
\noindent {\bf Acknowledgements}\\

\noindent This work was supported by NSF grant PHY-9507740. The authors 
wish to 
thank Lawrence Krauss for pointing out the relevance
of spatially flat models to small angular scale CMBR fluctuations.


\end{document}